# Role of Defect Induced Interfacial States in Molecular Sensing: Ultrahigh-Sensitive Region for Molecular Interaction

Rahul Tripathi[1], Pritam Bhattacharyya[2], Alok Shukla[2], Abha Misra[1]*

[1]Department of Instrumentation and Applied Physics, Indian Institute of Science, Bangalore, Karnataka, India 560012
[2]Department of Physics, Indian Institute of Technology Bombay, Powai, Mumbai, India 400076

**ABSTRACT**

The defect induced interfacial states are created in an atomically thin two-dimensional molybdenum disulfide channel by underlying a narrow pattern of a graphene layer in a field effect transistor. Nondestructive method for the generation of charge-state allowed a highly sensitive molecular interaction with the sensitivity of nearly three-order of magnitude at room temperature. The presence of interfacial states in the channel lead to a conductance fluctuation and its magnitude is modulated using the nitrogen dioxide gas molecules in the subthreshold region. The study provides a systematic approach to establish a correlation between modulated conductance fluctuation and the molecular concentration upto parts-per-billion. First-principles density functional theory further explains the role of unique interfacial configuration on conductance fluctuation. Therefore, our study demonstrates an experimental approach to induce charge-state for the modulation of carrier concentration and exploits the role of defect induced interfacial states in atomically thin interfaces for the molecular interaction.

**INTRODUCTION**

The defect states in oxide semiconductor interface play a crucial role in carrier transport across the channel in a field effect transistor (FET). Transistor characteristics such as charge carrier density, threshold voltage, device reliability etc. are greatly hindered by the defect states present in the interface or within the channel itself. A decoupling between vertical carrier transport (from bulk to the interface) and interface effects occurs in the presence of electrostatic field normal to the surface in case of bulk semiconductor.[1] Whereas in ultrathin semiconductors (0.7-5 nm), the carriers are transported along the interface to populate the channel in the presence of electrostatic field leading to a coupling between lateral carrier transport and interface properties. Recently, van der Waals (vdW) materials, such as n-type molybdenum disulfide ($MoS_2$) offer tunable bandgap (1-2 eV) with sub-nanometer thickness, high on-off ratios ($10^6 \sim 10^8$), efficient switching and extremely large surface to volume ratio.[2-4] These characteristics are highly desirable for tunable low power electronics like optoelectronic devices, memories, FET, light emitting devices and the gas sensors.[5-11] The uniqueness of $MoS_2$, in addition to field-effect properties, lies in its wide tunability of the field-dependent electronic states from localized to extended regimes and a well-controlled subthreshold properties. Intrinsic defects present in the $MoS_2$ channel alone, such as sulfur vacancy can seriously affect carrier transport properties in terms of mobility and subthreshold swing.[12,13] However, interface defects ($D_{inter}$) observed between depletion and accumulation regions in case of thin semiconductor can only be observed by flat-band voltage shift in the hysteresis measurements in contrast with the oxide defects.[14] A conductance fluctuation in the subthreshold region, generally observed at lower temperatures, can be explained in terms of electrostatic potential fluctuation in the channel due to the charged impurities at $SiO_2$ surface as well as in the MoS$_2$ as reported by Fang et al.[15] A static model is developed by Gaur et al to co-relate free and trapped charge densities with gate bias and can also be used to study the transport behavior and interface properties.[16] Furthermore, in the extraction of density of interface defects by conductance method has certain advantages over other capacitance method,[17] where these studies are mostly restricted to FET and frequency dependent applications only.[15-18]

Robinson et al. first observed a large electronic response in single-walled carbon nanotubes due to charge transfer at defect sites in the presence of analytes.[19] Recently, density functional theory predicted that MoS$_3$- and S-vacancy defective monolayer of MoS$_2$ has stronger chemisorption and greater electron transfer effects than pure MoS$_2$.[20] The presence of defects and point vacancies in the MoS$_2$ crystal structure facilitates the adsorption of different gas species, which strongly affect the transistor electrical characteristics.[21] Experimentally, the defects introduced by local irradiation of electron/ion beam significantly affect the properties of the MoS$_2$, leading to considerable changes in its electronic structure and transport behavior.[15,22,23] Therefore, there is severe requirement of developing nondestructive methods for the generation of defect states. Moreover, so far, no direct correlation is yet established between interfacial defects and modulated subthreshold transport in MoS$_2$ due to the molecular interaction.

In this work, the defect states are implanted in the channel of MoS$_2$ using a narrow pattern of a few-layer graphene (FLG) for the coupling of lateral carrier transport (in-plane channel resistance) and interface defects, $D_{inter}$. The defect induced interfacial states cause a conductance fluctuation in the sub-threshold region at room temperature with a high modulation using the nitrogen dioxide (NO$_2$) molecular concentration. Defect induced interfacial states near the conduction band leads to a movement of potential in the channel and causes a reduction in the number of free electrons in the channel. Therefore, an enormously large change in the current response upto ~$10^3$ was measured upon exposure to sub-parts-per-million (ppm) level of the molecular concentration. The molecular interaction is observed much higher compared to any previously reported results obtained with the pristine $MoS_2$ channel.[12-14,24-32] Thus, the artificially created defect states in vdW interfaces without uncontrolled doping or radiation damages paves an excellent route towards the elucidation of molecular interactions.

**RESULTS**

The defect induced interfacial states were verified using first-principles density functional theory (DFT) calculations. Figure 1(a) presents the computed band structure for the heterostructure of monolayer graphene and $MoS_2$ without any defect in the graphene layer. The band-structure and density of states (DOS) plot of the vdW heterointerface are presented in the Figure 1(b). The optimized geometry (details are provided in the supplementary information, Figure S1)[33] is shown in the inset of DOS profile in Figure 1(b). The band structure reveals the presence of both the conical type band structure of graphene and the band gap due to the semiconducting nature of $MoS_2$, which are consistent with the literature.[34] Defect in the heterostructure is created by removal of two carbon atoms from the graphene lattice as depicted in Figure 1(c). The band structure of the defected heterostructure in the absence of electric field is presented in Figure 1(d) revealing the presence of a defect state in the band structure. Thus, the absence of carbon atoms from the graphene in the $MoS_2$-graphene heterostructure has created an extra band near the Fermi level. The state contributed mainly from C-orbital plays a crucial role in the charge transport in the presence of external electric field. When an electric of -0.5 V/Å is applied in the perpendicular direction of the two-dimensional plane of the heterostructure, the gap remains the same and the number of bands near the Fermi level decreases as represented in Figure 1(e). However, at electric field of 0.5 eV/Å, it is observed that the defect induced interfacial state is well within the conduction band as well as the number of bands near the Fermi level increased significantly (Figure 1(f)). A band like transport is expected, which explains the metal-insulator transition in $MoS_2$.[35] The valance band minima shifts upwards at a negative external field leading to a reduced carrier concentration, thus, indicating the depletion region of the device characteristics.[35] As the field is positive, the Fermi level is at the vicinity of the interfacial states and begin to fill up. As a result, a conductance fluctuation is expected in the transport properties in the subthreshold region. The artificially created defect structure was computed in the absence of electric field and the DOS profile of the device is presented in the Figure 1(g). The DOS calculations were performed in both the absence and presence of external electric field which was applied perpendicularly to the plane of the system. We observed large change in the Fermi level in the DOS profile with the modulation in electric field. The change in DOS due to interfacial states is clearly observed in Figure 1(h) around 0.5 eV near conduction band minima. With a positive applied field, the state moves inside the conduction band and not visible in Figure 1(i). The state is prominent with negative applied field and can affect the subthreshold region where the DOS is relatively lower compared to positive electric field. Movement of the Fermi level towards more populated DOS with external positive field indicates more carrier density in the channel as shown in Figure 1(i). Furthermore, the band-structure of graphene/$MoS_2$ heterostructure with line defect in graphene lattice is also computed and shown in Figure S1(b-e) with and without presence of the electric field. It is also true that there is a significant difference in the band structure with the line defect; however, the physics still remains the same.

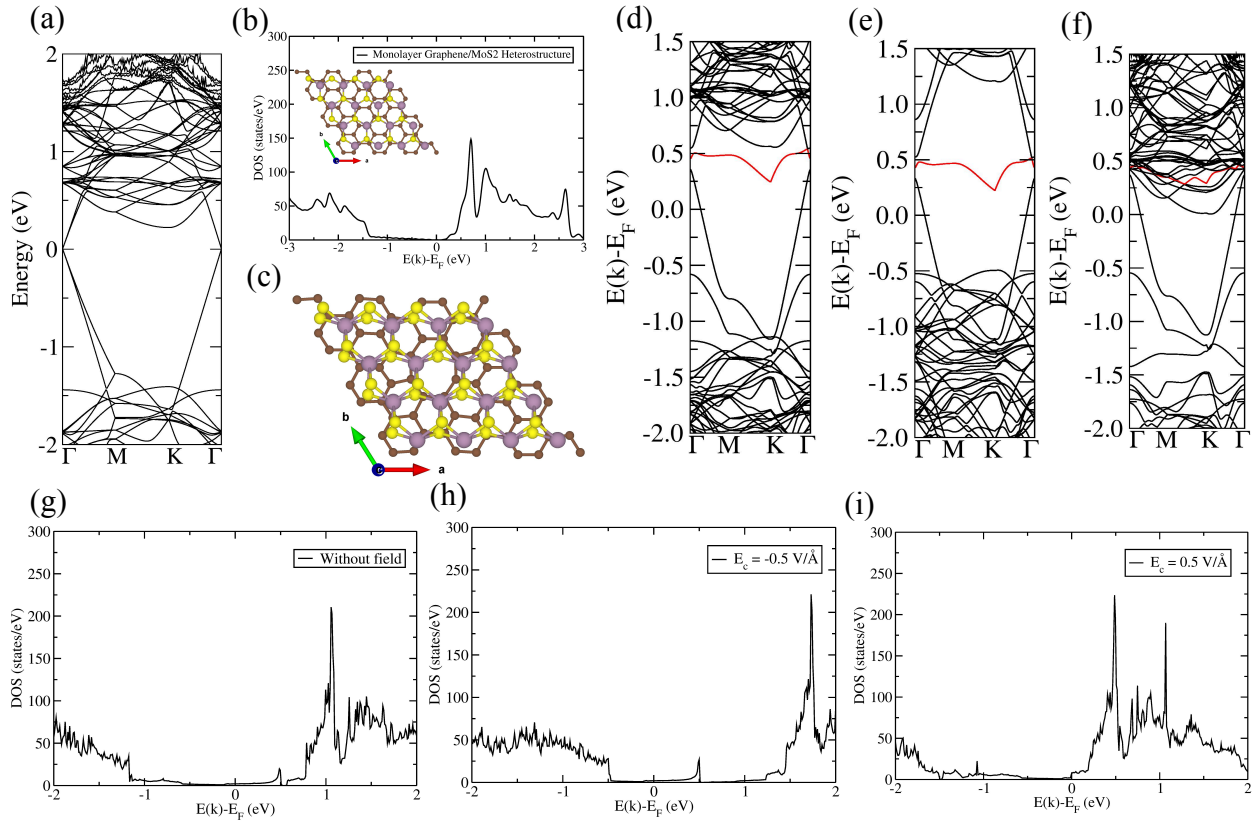

Figure 1. (a) Band structure and (b) DOS profile of Graphene/$MoS_2$ heterostructure computed in absence of external electric field, whereas inset of the DOS profile indicates its corresponding geometry. (c) The optimized geometry of the defected monolayer graphene/$MoS_2$ heterostructure, where two carbon atoms are missing from the graphene lattice. The band structures of the heterostructure in presence of (d) Zero, (e) -0.5, and (f) +0.5 V/Å electric field. The defect state is represented by red color in the band structures. (g) DOS profile of defect in Graphene/$MoS_2$ heterostructure without external field. (h) Movement of the interfacial state with negative and (i) positive external field

The interfaces of graphene and MoS$_2$ were obtained by exfoliation using the scotch tape method on the SiO$_2$ layer with 300 nm thickness deposited on a Si wafer. The pre-patterned metal contacts were fabricated using standard optical/electron beam lithography, followed by thermal deposition of Cr/Au contacts (10/20 nm). An FLG was then carefully transferred within the gap between the electrodes (not touching the drain electrodes) using a well-known polydimethylsiloxane dry transfer technique (experimental details are provided in the supplementary information Figure S2).[36,33] The transferred FLG was patterned with a gap of 1 μm using lithography and subsequent oxygen plasma reactive ion etching, which is the active MoS$_2$ channel in the device. The hetero-interface was formed by transferring exfoliated MoS$_2$ on the patterned FLG. The device design is depicted by a schematic in Figure 2(a) where the patterned FLG is the bottom layer whereas top MoS$_2$ layer alone forms a channel across the two metal electrodes. Back-gate voltage ($V_{bg}$) to the P$^{++}$ Si substrate controls the modulation of the charge carriers in the MoS$_2$ channel. The atomic force microscopic (AFM) image of the fabricated device is shown in Figure 2(a). The thickness of MoS$_2$ and graphene flakes are 4.5 nm and 5.8 nm, respectively. A few-layer thickness of both the graphene and MoS$_2$ was confirmed by Raman spectroscopy (Figure 2(b)). The Raman spectrum of MoS$_2$ depicts the characteristic $E^1_{2g}$ and out of plane $A_{2g}$ peaks at 384.7 and 407.1 cm$^{-1}$, respectively. Typical Raman signature of graphene flake is observed with G and 2D peaks.[37] The reported work function of pristine monolayer graphene is nearly 4.35 eV,[38,39] which is rather independent of the number of layers in FLG and converges to approximately 4.43 eV.[40] Moreover, FLG showed a p-type characteristic probably due to photoresist residue and charge impurities in the underlying substrate.[41] The role of oxygen molecules as well as moisture with graphene have also been reported to result in p-type dopant effects.[42] Although, water vapor does not dope graphene noticeably, yet it greatly impacts hole doping caused by oxygen molecules.[43] On the other hand, generally, $MoS_2$ shows n-type behavior due to S-vacancies and other lattice impurities with an electron affinity ($\chi_{MoS2}$) of ~ 4.15 eV.[44-47]

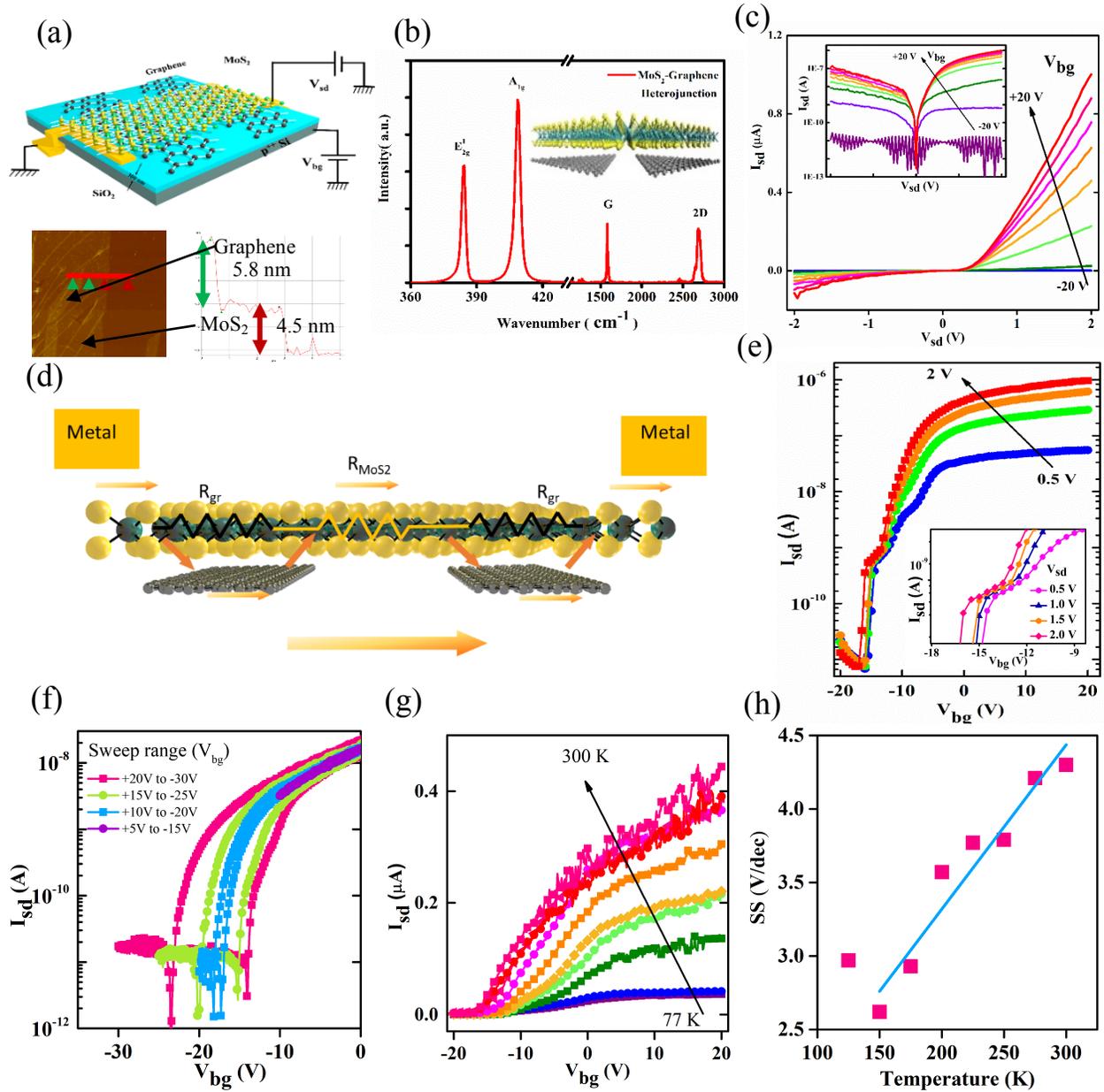

Figure 2. (a) Schematic of the device design. Below is the measured thickness of $MoS_2$ and graphene flakes using AFM. (b) Raman spectrum of the vdW interface. The inset shows perspective view of the device. (c) $I_{sd}$-$V_{sd}$ response in a range of −20 to +20 V, $V_{bg}$ showing a clear rectification behavior. The logarithmic plot of $I_{sd}$-$V_{sd}$ is shown in the inset. (d) The resistor network illustrates the role of different resistance component at the interface of $MoS_2$ and graphene. The orange arrows depict the flow of current path through the channel. (e) $I_{sd}$-$V_{bg}$ response for positive $V_{sd}$ bias where conduction fluctuation is observed. Inset shows the shift in the $V_{th}$ with the gate voltage. (f) Transfer characteristics of the device for different back-gate voltage in loops of different amplitudes but with fixed steps ($\Delta V_{bg}$=0.1 V) at room temperature. (g)Temperature dependent drain current variation of the device at low bias. (h) Linear dependency of sub-threshold swing at different temperatures.

Electrical measurements were performed at both room and low temperatures for both the devices (D1 and D2) in a controlled environment. Figure 2(c) demonstrates $I_{sd}$-$V_{sd}$ response of the vdW interface with defect states, showing a rectification behavior with $V_{bg}$, expected for a heterojunction at the interface of $MoS_2$ and graphene at room temperature. The observed rectification behavior is qualitatively like a conventional *p-n* junction, although the current does not increase exponentially with the applied forward bias. Linear increment in the current and the underlying rectification mechanism can be attributed to the barrier formation between $MoS_2$-graphene and $MoS_2$-metal contacts and the recombination process at the atomically thin interface.[48,49] Furthermore, the asymmetric nature of the drain current along with the recombination process can be explained by two back-to-back Schottky barriers.[50] A slight difference in geometrical contact area plays a crucial role in the symmetric nature of *I-V* curve.

The asymmetric nature observed in $I_{sd}$-$V_{sd}$ curve with different $V_{bg}$ is possibly due to the different barrier heights in source drain contacts.[50] The S-D bias of any polarity applied on the device sets one Schottky barrier in the reverse direction and the other in the forward direction. Hence, the drain current is always limited by a reverse biased junction in the device. Inset in Figure 2(c) demonstrates that the drain current is well modulated by $V_{bg}$, although there is a less dependency on $V_{bg}$ towards the negative drain bias region (Figure S3(a))[33]. The total channel resistance ($R_{CH}$) can be seen as a combined resistances of both graphene ($R_{gr}$) and the channel $MoS_2$ ($R_{MoS2}$). The effective channel length plays a critical role as the channel resistance ($R_{CH}$) dominates with the large channel length compared to the defect induced resistance as predicted by Gaur et al.[16] The direction of current flow (orange arrow) in the channel is shown in Figure 2(d) where the role of artificially induced interfacial defects play a significant role in the flow of charge carriers. Density of defect state can be measured easily in the small channel length, where the free carrier concentration is low and the defect induced interfacial states can be easily tuned (filled or unfilled) by $V_{bg}$ in $MoS_2$.[51] The fluctuation in electrostatic potential of the channel due to charged impurities at the $MoS_2$-graphene interface can be reflected in lower carrier concentration region in terms of channel resistance. The channel resistance can be seen as a combined effect of free carrier density and mobility. Mobility in the subthreshold region is dominated by charge impurity scattering of free carriers by the charged defect states.[17] Interface defects are observed between depletion and accumulation regions. As the potential increases, these charged defect states are screened by the free carriers and mobility increases. In accumulation (ON-state) region, mobility can be described by the trap and release of carriers by the interfacial states. This reduces the effective mobility of the carriers. Combining these two effects, it is evident that in the presence of defect states, $R_{CH}$ is a function of $D_{inter}$ through carrier density and mobility. The transfer curve of the device plotted on a log scale in Figure 2(e) shows that an on-off ratio of the device is ~$10^5$ with the gate voltage variation from –20 to +20 V and the device exhibits n-type behavior with a threshold voltage ($V_{th}$) of around –15 V (for details see Figure S3(b)).[33] As mentioned earlier, the defect states present at the interface impact the subthreshold slope. Variation in the drain current is plotted in Figure 2(e) with back-gate voltage for only positive drain bias to illustrate a current fluctuation in the subthreshold region, which is highly prominent compared to negative source-drain bias (Figure S3(c)).[33] Defect states present in the interface immobilize the carriers and reduce the free carrier concentration in the majority band as can be seen from the intermediate constant current state. A poor subthreshold swing (SS) was observed in the subthreshold region around –14 V of back gate voltage. Furthermore, the inset in Figure 2(e) depicts a close-up view of the drain current in subthreshold region to reveal that an almost constant drain current is observed for a span of 3 V back-gate voltage with a slight broadening (0.5 V, $V_{bg}$) at the higher source-drain bias. A specific defect in the $MoS_2$ channel or the defect states at the interface of $MoS_2$-graphene may contribute to the conductance fluctuation. This conductance fluctuation is consistent with same measurement parameters measured for both the devices (D1 and D2), which is shown in the supplementary information. Moreover, a shift in the $V_{th}$ is also observed with the increase in the bias voltage (Figure S3(d)).[33] The presence of interfacial defect states at room temperature was further confirmed by the hysteresis measurement of the device as shown in Figure 2(f). The voltage was swept from the positive $V_{bg}$ towards zero $V_{bg}$ and then reversed to the negative $V_{bg}$ values and finally back to +20 V of $V_{bg}$. For each successive curve, the gate voltage range was increased by 10 V. The larger hysteresis loop was observed for $V_{bg}$ sweep from +20 to –30 V, which corresponds to an effective trapping/detrapping of the trapped charges due to the presence of defect states.[52] The width of hysteresis loop was reduced with the less sweeping range of $V_{bg}$ and completely disappeared for the sweeping range from +5 to –10 V of $V_{bg}$. The field effect mobility can be obtained from the transfer characteristic curve for a fixed bias using the following relation, $\mu_{FE} = \frac{L}{WC_{OX}V_{sd}} \frac{dI_{sd}}{dV_{bg}}$, where $L$ and $W$ are the device channel length and width, respectively, $C_{ox}$ is the oxide capacitance per unit area of 300 nm thick $SiO_2$ layer, $V_{sd}$ is the applied drain bias and $\left(\frac{dI_{sd}}{dV_{bg}}\right)$ is the slope of the curve. The field effect mobility of the device at room temperature is calculated to be 1.7066 $cm^2V^{-1}s^{-1}$ with a subthreshold swing of 4.3V/dec. The performance of the back-gated interface of $MoS_2$-graphene FET device under different temperature conditions (77–300 K) was investigated in Figure 2(f) at a low bias of 100 mV at different back-gate voltages. The SS-factor is closely related to the temperature due to the defect states present at the interface according to the conventional semiconductor theory.[53] The SS is plotted with the temperature in Figure 2(g) which shows a linear increase with the temperature. The defect induced interface state density can be extracted from the slope as the contribution is negligible at depletion layer due to the gate capacitance owing to the atomically small $MoS_2$ channel thickness. Therefore, the density of defect induced interface state ($D_{inter}$) can be estimated from the temperature dependence of the SS-factor by using the equation $D_{inter} = \frac{C_{bg}}{e}\left(\frac{dS}{dT}\frac{e}{k_B}\frac{1}{\ln 10} - 1\right)$, where $k_B, e, C_{bg}$ is Boltzmann constant, charge of electron and the geometrical capacitance, respectively. The calculated $D_{inter}$ is $4.22 \times 10^{13}\ cm^{-2}eV^{-1}$, which shows that an acceptor type high density of defect sates that are present near the conduction band minima, primarily caused by the combination of interface defects and grain boundary dislocations.[17] The concentration of trap charges due to defect states can be easily found by integrating $D_{inter}(E)$ for all the available energy levels between conductance band and valance band. We estimated the charge density to be in the order of $10^{13}/cm^{-2}$ per unit energy level.

The effect of defect induced interfacial states in molecular sensing ability of vdW interface was evaluated in $NO_2$ gas environment. Nitrogen was used as a carrier gas to dilute the target gas molecules to the desired concentration through the flow-rate adjustment using mass flow controllers. The $I_{sd} - V_{sd}$ response for different molecular concentrations (100 ppb – 5 ppm) is shown in Figure 3(a). A dramatic device response was measured in the charge transport characteristics upon exposure to the molecules. There is a large decrease (~200%) in the current at +20 V, $V_{bg}$ after the gas exposure to the device due to its electron withdrawing nature.

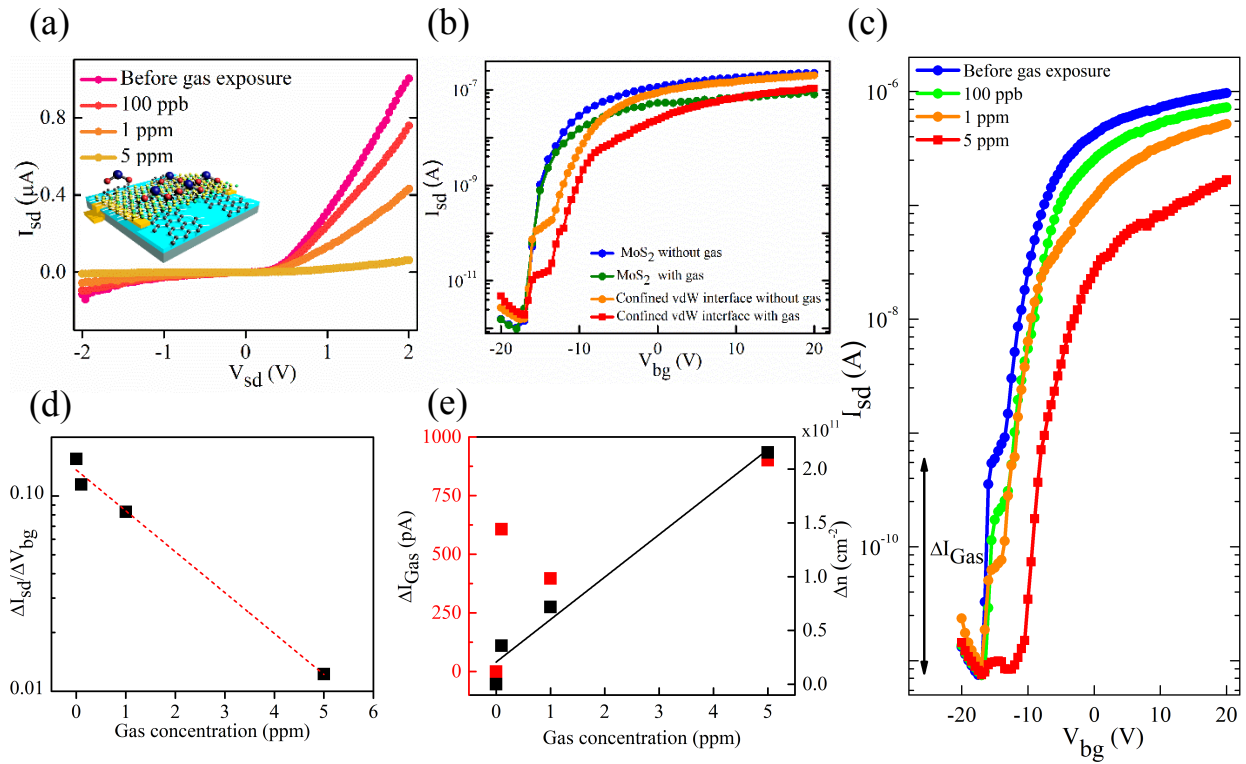

Figure 3. (a) Evaluation of $I_{sd}$-$V_{sd}$ response with different $NO_2$ gas concentrations. (b) Comparison

of the $I_{sd}$-$V_{bg}$ response of the pristine MoS$_2$ device with the defect induced vdW interfaces at a fixed gas concentration of 100 ppb. (c) The stepwise decrement of current in the subthreshold region with the molecular exposure. (d) The systematic variation in $\Delta I_{sd}/\Delta V_{bg}$ with the different concentrations of molecular exposure in the subthreshold region. (e) Change in the drain current and the carrier density modulation in the subthreshold region due to molecular exposure.

Figure 3(b) shows the evaluation of gate response of the device at a constant drain bias ($V_{sd}$ = +2 V) towards the exposure of 100 ppb NO$_2$. The real-time molecular interaction response of the device is compared with the pristine MoS$_2$ based FET. Interestingly a large change of 1000% (~10 times) in the current is observed in the subthreshold region (−14.6 V, $V_{bg}$) upon exposure to the molecules. A large decrease in the current response is observed in the subthreshold region as compared to the depletion region or accumulation region as also discussed later in figure 4. Thus, it is clear from both the results in figure 3(a) and (b) that the molecular response of the device is highly dependent on the gas concentration. Figure 3(c) illustrates that the device response systematically decreases in the subthreshold region upon increasing the concentration of gas molecules from 100 ppb to 5 ppm. In the subthreshold region, the change in the current due to the gas exposure, ($\Delta I_{Gas}$) is almost three orders of magnitude higher at 5 ppm concentration of gas. This is an extraordinarily large change reported so far with respect to previously reported results.[12-14, 31-36] The change in slope is observed with the variation in the gate voltage, $\Delta I_{sd}/\Delta V_{bg}$ as depicted in figure 3(d), which is found extremely sensitive to the molecular interaction. A decay in $\Delta I_{sd}/\Delta V_{bg}$ was noted with the increase in the molecular concentration indicating the gas concentration dependent restricted movement of the electrons. It decays to 31.4% in the presence of 100 ppb of gas exposure. At 5 ppm of gas concentration, the value of $\Delta I_{sd}/\Delta V_{bg}$ becomes nearly zero, *i.e.* completely switching off the channel (Figure S4(a)).[33] Furthermore, the vertical shift in current ($\Delta I_{gas}$) in subthreshold region at −14.6 V, $V_{bg}$ under molecular exposure is also investigated at ppm and sub-ppm levels. The change in current follows a linear relationship with molecular concentration in case of +20 V region ((Figure S4(b)).[33] Thus, the device behaves as a molecular switch and completely switches off from its on-state in the presence of gas molecules for all the devices (Figure S4(c) and (d)).[33] The change in carrier density with molecular exposure in the channel is also plotted in Figure 3(e) showing a linear dependency with molecular concentration. The change in carrier density due to molecular exposure is found out to be in the order of $10^{11}/cm^{-2}$ which is much lower than charge density ($10^{13}/cm^{-2}$) due to $D_{inter}$ per unit energy level as calculated earlier. So, the defect induced interfacial states play a crucial role in transport properties in the presence of external molecules, where the carrier concentration is low in the MoS$_2$ channel.

Although in many reports so far, the operational region for gas sensing is the accumulation region (+20V, $V_{bg}$),[10-11] but the subthreshold region becomes an extremely sensitive region of the device because of the induced defect states in the channel at MoS$_2$ -graphene interface. Figure 4(a) depicts the analysis of the current obtained from Figure 3(b) at $V_{bg}$ = −18, −14 and +10 V for depletion, subthreshold and accumulation regions, respectively at both different drain bias and different molecular concentration.

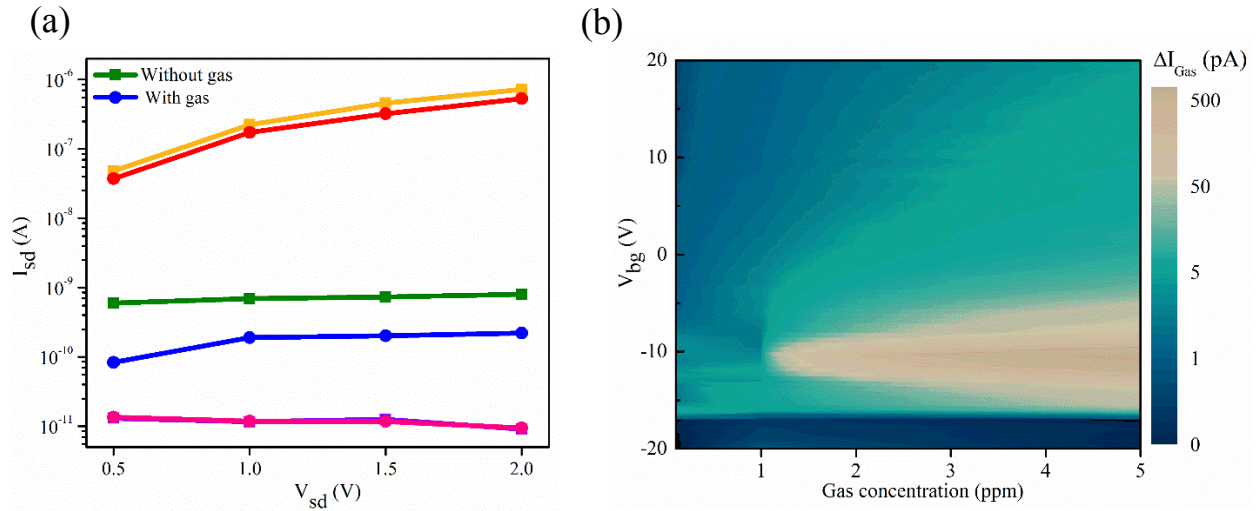

Figure 4. (a) Evolution of current response in depletion, subthreshold and accumulation regions with NO$_2$ gas interaction at 100 ppb concentration. (b) Mapping of the regions depicting the change in the current ($ΔI_{Gas}$) with gas interaction.

This clearly implicates that the response of the device is solely depend on the gate bias rather than the source-drain bias which is observed to be more in the subthreshold region. For the gas concentration of 100 ppb there is a large decrease in the current value (~1000%) in the subthreshold as compared accumulation or depletion regions because of the restricted movement of the electrons at the interface with the applied external electric field. On other hand, overall modulation in the drain current was also mapped in presence of different molecular concentration with different gate voltages showing a conventional charge transfer process by the molecules which can be verified by a simple classical model (FigureS5). The change in drain current with molecular interaction was further evaluated by mapping of $ΔI_{Gas}$ as depicted in Figure 4(b). The mapping shows the response due to molecules is mostly observed in the sub-threshold region and broadened more with the increasing molecular concentration. So, the defect induced interfacial states are more sensitive towards molecules at lower carrier concentration region. The width of only MoS$_2$ channel in our case is 2 microns, which is considered crucial to observe the role of defect states present in the channel.[16] Furthermore, the carrier injection process from n-type MoS$_2$ to a p-type graphene or vice versa is important where the interface plays a critical role in the charge transport. Subthreshold slope and the temperature dependent measurements show a clear dependency on defect states, which clarify that the MoS$_2$ in the middle plays a crucial role in the transport properties. The role of MoS$_2$ and the presence of defect states in the MoS$_2$ can be further studied using a spatial variation width of MoS$_2$ channel. It would be an interesting fact that the intrinsic defects present in the MoS$_2$ channel can interact with the external molecules and can be seen in transport properties at the mesoscopic scale.

**CONCLUSION**

The patterned structure of graphene in MoS$_2$-grpahene vdW interface leads to a defect induced interfacial states for engineered molecular interaction using much lower concentration of NO$_2$ gas molecules. The molecular interaction through the defect induced interfacial states is observed for a deep insight into the atomically designed interface. An anomalous change of three orders of magnitude in the drain current was observed in the sub-threshold region by the molecular interaction at the ppb level concentration of the molecules. This anomalous behavior of the device is explained using a classical model calculation and DFT simulation. DFT simulation shows that the removal of carbon atoms from the graphene modulates the DOS in the MoS$_2$-graphene system, which plays a crucial role in the electrical transport in the device. Furthermore, the defect induced interfacial states slows down the movement of potential in the channel, prominent in lower carrier concentration region in the channel leads to a restricted movement of electrons that has a strong dependence on the molecular concentration as like a molecular switch. Therefore, our investigation entices a new route to design and construct heterointerfaces for new generation platform to study molecular interactions.

**MATERIALS AND METHODS**

**Device fabrication**

We followed a standard procedure of dry transfer technique using PDMS stamp without using any sacrificial layer to ensure the good surface quality. A piece of glass was covered with a PDMS stamp (0.2-0.4 mm thick) which is transparent. Few layers of graphene (HQ Graphene) were then exfoliated directly on the stamp and examine under a microscope to choose the proper flake. Contacts were drawn in a $Si/SiO_2$ substrate using standard lithography process, and metallization was carried out with Cr/Au (5 nm/40 nm). The chosen graphene flake was then carefully transferred between the source-drain (S-D) electrode and oxygen plasma etching was done to pattern the graphene into the desired shape. Now, few layers of the $MoS_2$ flake (Manchester Nanomaterial) was exfoliated onto the PDMS stamp and placed on the micromanipulator stage upside down and aligned with the patterned graphene flake using the microscope. The two were bought into contact and $MoS_2$ flake was transferred on top of graphene.

**Electrical and gas sensing measurements**

Raman spectroscopy was performed in the air after back-gate transistor fabrication, at a wavelength of 532 nm and with a light spot of 1.1 microns. The electrical characterization of the devices was carried out with an Agilent B1500A semiconductor analyzer in a closed chamber with a nitrogen environment. Temperature dependent measurement was done in a vacuum chamber using Agilent B1500A semiconductor analyzer with a ramp of 25ºC. All the gas sensing experiments were performed at room temperature and atmospheric pressure. The gas sensing measurement was done by mixing the gas with pure nitrogen to achieve desirable concentration before being injected into the chamber. Mass flow controllers were used to control the concentration, and the total flow rate was 1000 standard cubic centimeters per minute for all time. The gas inlet was designed to point directly onto the sample with a gap of ~5 mm for the quick response.

**Theoretical Approach and Computational Details**

We have performed the first principles study within the framework of density functional theory[54] using the computer program, Vienna Ab-initio Simulation Package (VASP).[55,56] All the calculations were carried out using projector-augmented wave (PAW) pseudo-potentials[55,57] and Perdew-Burke-Ernzerhof (PBE) exchange correlation functional.[58] 500 eV was chosen as the kinetic energy cut-off for plain wave basis set. First, the monolayer hetero structure was generated using the CellMatch code.[59] Next, the defects in the heterostructure were created by removing two carbon atoms from its middle portion, and the defected heterostructure was optimized using a k-point mesh of 3×3×1. The energy convergence criteria for electronic steps were set $10^{-5}$ eV, while the convergence of Hellmann-Feynman force for the ionic steps were $10^{-2}$ eV/Å. A k-point mesh of 15×15×1 was employed to perform the density of state (DOS) calculations. For the purpose, we set $10^{-6}$ eV as the threshold of energy convergence and included vacuum of 18 Å along the perpendicular direction of the 2-dimensional (2D) plane of the heterostructure to reduce the interactions with its images. In this study, van der Waals force correction was included by employing Grimme's DFT-D2 method in our calculations.[60] The band-structure calculations of the defected heterostructure were performed using 60 k-points in the reciprocal space.